# Photoluminescence Path Bifurcations by Spin Flip in Two-Dimensional CrPS$_4$

Suhyeon Kim[1#], Sangho Yoon[2,3#], Hyobin Ahn[4], Gangtae Jin[2,3], Hyesun Kim[1], Moon-Ho Jo[2,3], Changgu Lee[4], Jonghwan Kim[2,3*], and Sunmin Ryu[1,5*]

[1]Department of Chemistry, Pohang University of Science and Technology (POSTECH), Pohang 37673, Korea

[2]Department of Materials Science and Engineering, Pohang University of Science and Technology (POSTECH), Pohang 37673, Korea

[3]Center for Van der Waals Quantum Solids, Institute for Basic Science (IBS), Pohang, 37673, Korea

[4]School of Mechanical Engineering, Sungkyunkwan University, Suwon 16419, Korea

[5]Institute for Convergence Research and Education in Advanced Technology (I-CREATE), Yonsei University, Seoul 03722, Korea

**ABSTRACT**

Ultrathin layered crystals of coordinated chromium(III) are promising not only as two-dimensional (2D) magnets but also as 2D near-infrared (NIR) emitters owing to long-range spin correlation and efficient transition between high and low-spin excited states of Cr$^{3+}$ ions. In this study, we report on dual-band NIR photoluminescence (PL) of CrPS$_4$ and show that its excitonic emission bifurcates into fluorescence and phosphorescence depending on thickness, temperature and defect density. In addition to the spectral branching, the biexponential decay of PL transients, also affected by the three factors, could be well described within a three-level kinetic model for Cr(III). In essence, the PL bifurcations are governed by activated reverse intersystem crossing from the low to high-spin states, and the transition barrier becomes lower for thinner 2D samples because of surface-localized defects. Our findings can be generalized to 2D solids of coordinated metals and will be valuable in realizing groundbreaking magneto-optic functions and devices.

Electronic spin is an important degree of freedom that governs the material properties of various two-dimensional (2D) crystals. The frontier electronic bands of transition metal dichalcogenides (TMD) represented by $MoS_2$ and $WS_2$ are split by spin-orbit coupling[1-3] and can be selectively photoexcited using circularly polarized light.[4] Spins in some 2D Cr-based crystals, including $Cr_2Ge_2Te_6$ and $CrI_3$, are coupled by exchange interaction and exhibit ferromagnetism or antiferromagnetism depending on their interlayer interactions.[5, 6] The spin states of 2D materials have also been targeted for exploitation to store and transport digital information, which evolves into graphene spintronics[7, 8] However, the spin degree of freedom has not been probed for its role in the excited-state dynamics of 2D materials except recent studies on valley pseudospin in TMDs.[9-11] Viewing that the fate of photogenerated excitons in molecular systems is complicated and branched by multiple electronic states with different spin multiplicities,[12] 2D material's excitonic dynamics involving radiative and nonradiative transitions may well be enriched by the spin. Understanding their relaxational pathways is crucial in harnessing absorbed energy in photovoltaics[13] and photocatalysis,[14] devising efficient photonic applications,[15] and mitigating photoinduced self-degradation.[16]

In this regard, metal (M) phosphorus sulfides ($MPS_3$ and $MPS_4$) and trihalides ($MX_3$; X = Cl, Br or I)[17] with M under octahedral coordination[18] can serve as a good model system to study spin-dependent exciton behaviors. Their lowest electronic transition is mostly of d-d character and localized at the transition metal atoms.[17, 19-21] This simplifies the description of the excited states, which stands in contrast to the case of TMDs with strong M-chalcogen hybridization in their frontier electronic bands.[1-3] The same fact also indicates that the primary excitonic behavior is governed by the spin configuration of M and consequently the field strength of the ligand species coordinated to M.[22] This further suggests the possibility that the excitonic dynamics can be tuned by varying the ligands.[23] Lastly, they are layered magnetic materials and can be prepared in 2D forms serving as monolayer magnets.[5, 6] It may allow the manipulation of excitonic dynamics via controlling the spin-spin exchange interactions. Among the aforementioned magnetic semiconductors, chromium thiophosphate ($CrPS_4$) can be a prime example. Widely varying photophysics of $Cr^{3+}$ ions, as such in $CrPS_4$, have been documented for various ionic complexes[24, 25] and dopant-host systems[26] including ruby used as the first laser's gain medium.[27] Single and few-layer $CrPS_4$ exhibiting distinctive absorption and emission attributed to $Cr^{3+}$ ions[21, 28] possesses highly anisotropic optical response[29] and lattice dynamics.[30] It also hosts quantum interference in photoluminescence (PL) attributed to Fano resonance[31] and allows sensitive NIR detection,[32] possibly with polarization control.[33] Despite the photochemical oxidation limiting its ambient and photo-stability,[34] ferromagnetism has recently been demonstrated for single-layer $CrPS_4$.[35]

In this work, we report on the temperature-dependent PL bifurcations of 2D and bulk $CrPS_4$. The NIR PL signals originating from ligand-perturbed $Cr^{3+}$ ions ($^4A_{2g}$ in the ground state) are entangled by highly efficient intersystem crossing (ISC) between two excited ionic states ($^4T_{2g}$ and $^2E_g$) of different spin multiplicity. Whereas the phosphorescence from the lower doublet state is dominant below 100 K, it is mostly replaced with the fluorescence from the upper quartet state thermally populated by reverse intersystem crossing (RISC) from the doublet state at higher temperatures. The relaxation of the Cr-localized excitons is accelerated by the activated radiative transition at elevated temperatures and defect-induced quenching preferred by thinner 2D forms. The bifurcated excitonic emissions from 2D $CrPS_4$ will be useful in the diagnosis of electronic structures and magneto-optoelectronic applications.

**Results and Discussion**

***Dual-band PL and its substantial Stokes shift.*** 2D and bulk-like thick CrPS$_4$ samples were prepared by mechanical exfoliation, and their surfaces were encapsulated with Al$_2$O$_3$ films for protection against oxidation as described in Methods. In Fig. 1a, we compare their PL spectra obtained respectively with the visible (VIS) and NIR detectors (see Methods) and show that the PL band of CrPS$_4$ is broader, more structured and located at lower energy than previously reported.[29, 31] The VIS PL band of 4L centered at 1.33 eV almost coincides with that of bulk, which verifies the negligible dependence of exciton energy on thickness.[29] Whereas the NIR bands show the same tendency, they were found ~200 meV downshifted compared to the VIS bands. As shown in Fig. S1, the apparent shift is due to the line-shape distortion caused by the limited spectral sensitivity of the Si-based VIS detector below 1.3 eV. This fact explains why the lower energy emission was not recorded in the previous papers[29, 31] and indicates that the PL signals detected by VIS detectors are only a minor subset of the whole. With the lowest absorption peak at ~1.7 eV,[29] the PL peak energy of 1.13 eV revised in this work gives a large Stokes shift of 0.57 eV. This finding implies that a substantial nuclear relaxation occurs in the excited states, as discussed below.

The structured NIR spectra indicate that the emission contains two bands possibly divided by the kink at ~1.2 eV. As described in earlier reports,[29] the low-energy absorption and PL of CrPS$_4$ can be approximated as the atomic transitions of Cr$^{3+}$ ions under octahedral ligand-field (LF) perturbation. As depicted in Fig. 1b, the monoclinic crystal of CrPS$_4$ belonging to the space group $C_2^3$ is layered with a interplanar distance of 6.137 Å and its unit cell spans with ***a*** = 10.871 Å, ***b*** = 7.254 Å and ***c*** = 6.140 Å (β = 91.88°). Each Cr atom is coordinated by 6 S atoms, and the resulting CrS$_6$ octahedra are aligned along the ***b*** axis, forming quasi-1D structures. As the octahedral perturbation splits the d orbital of Cr$^{3+}$ ions into e$_g$ and t$_{2g}$ orbitals, the ground state ($^4$A$_{2g}$) and three lowest electronic states ($^2$E$_g$, $^4$T$_{2g}$ and $^4$T$_{1g}$) emerge from the ground term ($^4$F) of d$^3$.[22] Note that the slight distortion in the octahedron would need to be considered for accurate symmetry analysis. Whereas the energy ordering is dictated by the LF splitting (Δ$_o$) according to the Tanabe-Sugano diagram in Fig. 1c, the doublet excited state is located near $^4$T$_{2g}$ in many octahedral Cr$^{3+}$ systems.[24] In fact, these two states turned out to be responsible for the dual-band PL as will be explained below.

We also note that a more accurate description of PL behavior requires potential energy surfaces in configurational coordinates that represent the metal-ligand separation. Figure 1c depicts the three representative cases proposed for various material systems of octahedral $Cr^{3+}$ ions.[26] As $^4A_{2g}$ and $^2E_g$ have the same electronic configuration ($t_{2g}^3$) and thus identical spatial wave function, their equilibrium coordinates are very close. In contrast, the equilibrium separation of the quartet excited state ($^4T_{2g}$) with a different configuration ($t_{2g}^2 e_g^1$) is substantially displaced from that of the ground state. After the photoabsorption by the spin-allowed $^4A_{2g} \rightarrow {}^4T_{2g}$ transition (black vertical arrow), the excited ions may follow a few different relaxation pathways depending on the energetic ordering of the two excited potential surfaces.[26] When the quartet state is lower than the doublet, radiative decay giving fluorescence (green arrow) will occur following a rapid vibrational relaxation (VR) within the quartet's surface, as shown in Fig. 1c (left). For the opposite ordering, the quartet state is converted to the doublet through an ultrafast ISC, which competes with the VR process.[36, 37] Then, the PL signals will be dominated by phosphorescence (purple arrow) from the spin-forbidden transition. In the degenerate case in Fig. 1c (middle), both emissions can be detected simultaneously but at different energies because of the large configurational displacement of $^4T_{2g}$.[24, 38]

*Temperature-dependent bifurcations of PL.* To disentangle the unresolved PL bands, we performed variable-temperature PL measurements using the NIR detector. By lowering the temperature to 100 K, the broad PL band (denoted as **F**) at 1.13 eV almost disappeared, and a multi-peak band (denoted as **P**) emerged at ~1.3 eV (Fig. 2a). With decreasing temperature further down to 50 K, the latter band became more dominant with clearer separation between neighboring sub-peaks. Below 50 K, another broad band (denoted as **D**) centered at 1.08 eV appeared, borrowing the intensity of **P** (see Fig. S2 for complete spectral analysis). Few-layer samples exhibited similar spectral changes with weaker **P** and stronger **D** (Fig. S3). The temperature-dependent switching between **F** and **P** indicates a thermally activated transition. As a modification to the schemes in Fig. 1c, we propose an intermediate-field arrangement with a low-spin ground state for $CrPS_4$ in Fig. 2b. The Franck-Condon transition of $^4A_{2g} \rightarrow {}^4T_{2g}$ corresponds to the lowest absorption peak at 1.7 eV.[29] The efficient ISC to the doublet surface minimizes the system's energy further than VR within $^4T_{2g}$, leading to the phosphorescence-only band (**P**) at 50 ~ 100 K. At higher temperatures, however, the ions in the $^2E_g$ surface can be activated to overcome the energy barrier between the two excited surfaces via RISC.[24] The ions now in $^4T_{2g}$ will decay by generating the fluorescence band (**F**), repeating ISC to $^2E_g$ or undergoing nonradiative relaxation. Despite the endothermicity of RISC, however, the energy of **F** can be smaller than that of **P** in the proposed scheme (Fig. 2b & 2c), which nicely explains what was observed in Fig. 2a.

We note that the changes in the line shape of **P** contain important information about excited CrPS$_4$. The fact that **P** is negligible above 150 K suggests that the energy barrier to RISC is on the order of 10 meV, suggesting that the two excited states are almost degenerate. Assuming that its sub-peaks originate from vibrational progression, it can be inferred that the energy minimum of the $^2E_g$ surface is significantly displaced from the ground state along another normal coordinate other than the one shown in Fig. 2b.[39] The average spacing (~32 meV) of the vibronic peaks suggests that the normal mode B$_7$ at 256.6 cm$^{-1}$ can be associated with the vibrational progression.[30] Upon decreasing the temperature from 298 to 4 K, the 0-0 peak energy of **P** increased only by 5 meV, which stands in stark contrast to 80 meV for MoS$_2$.[40] This corroborates our assumption that the PL transitions are largely localized within Cr$^{3+}$ ions and thus hardly affected by thermal expansion of lattice and electron-lattice interaction. Lastly, the attenuating **P** accompanied by the emerging **D** below 50 K suggests that the ions in $^2E_g$ relax to another state with a small energy difference. Note that the intensity of **D** is significant only below 16 K. We attribute the **D** state (denoted as a dotted line in Fig. 2b) to structural defects as found for 2D TMDs.[40] Indeed, **D** was more intense for 4L than bulk (Fig. S3), which is consistent with the fact that the former was more defective (shown later). Moreover, the bulk samples grown by chemical vapor transport (CVT) exhibited stronger **D** than those grown by the self-flux method,[41] as shown in Fig. S4. This fact suggests that the latter method generates better CrPS$_4$ crystals than the former. Whereas the structural nature of the defects is currently unknown, the **D** state is located slightly lower than $^2E_g$ and activated to the doublet state above 16 K. We will discuss other roles of defects in the excitonic decay of CrPS$_4$ later.

We also show that the temperature-governed excitonic bifurcations substantially influence the PL lifetime. For this, we measured time-resolved PL intensities employing the method of time-correlated single-photon counting (TCSPC) as given in Fig. 3a for 6L CrPS$_4$. Because the TCSPC setup was based on a VIS avalanche photodiode combined with a set of filters allowing transmission in the range of 1.24 ~1.55 eV, we first verified the PL spectra obtained with the VIS-CCD setup. As shown in Fig. 3a, a few vibronic peaks of **P** could be observed below 100 K with noticeable truncation in intensity below 1.3 eV. The bulk spectra provided a clearer separation of the vibronic peaks (Fig. S5a). Then, the signals in Fig. 3a, when filtered with the passband mentioned above, mostly originate from **P** and thus are phosphorescence. Clearly, the decay of PL signals is faster at higher temperatures for 6L (Fig. 3b) and bulk (Fig. S5b) samples. The enhanced relaxation can be attributed to the increase in the RISC rate at elevated temperatures (Fig. 2c). The analysis with a biexponential function led to a pair of time constants, $\tau_1 = 70$ and $\tau_2 = 390$ ps for 298 K, each of which increased twice for 77 K (see Supporting Note for a kinetic model). Notably, bulk samples exhibited a $\tau_2$-dominated decay that was 3 ~ 4 times slower than 6L at 298 K (Fig. S5c). This finding implies that the faster component insignificant in bulk samples is governed by one factor more influential in the 2D limit, which will be discussed below.

*Defect-modulated bifurcations in 2D CrPS$_4$*. We found that the excitonic decay CrPS$_4$ exhibits a distinctive dependence on its thickness. The PL transients in Fig. S6 remained unchanged for 250 ~ 7L but exhibited accelerated decay for thinner samples. The change was in the same direction as that induced by higher temperatures and yet more pronounced. As shown in Fig. S6, the transients could be well fit with a double-exponential function for all samples. Notably, detailed analysis showed that the overall PL behaviors divide at ~7L. Whereas the contribution of the fast component with a time constant $\tau_1$ was larger than the slow one for < 6L, it decreased rapidly and became insignificant with increasing thickness (Fig. 4a). The slow time constant ($\tau_2$) exhibiting the starkest contrast across the divide drastically varied in the first 7L and remained nearly constant up to the maximum thickness. Both time constants decreased with decreasing thickness. The PL transients were hardly affected by the excitation fluence (Fig. S7). In Fig. 4b, the effective quantum yield (QY$_{eff}$) of PL collected by the VIS-CCD, thus mostly phosphorescence, was determined under the assumption that the degree of absorption is proportional to the differential reflectance (DR).[42] When referenced to 7L, QY$_{eff}$ dropped by ~70% for samples < 6L, which agrees with the change in the PL lifetime (Fig. 4a).

We conclude that the unusual behavior of the thinner samples is induced by structural defects. Because the surface layers of layered crystals tend to have more defects than inner ones, the influence of defects becomes more significant with decreasing thickness.[43] Less defective bulk samples grown by the self-flux method exhibited a twice longer lifetime (Fig. 4a) and less **D** band (Fig. S4) than those by CVT. As a control test, artificial defects were generated by exposing 6L samples to UV-generated ozone: the PL spectra decreased in intensity but without a noticeable change in their line shape (Fig. 5a). It is to be noted that an exposure of 2 min reduced the PL intensity by 45%, although it did not cause any significant change in Raman spectra (Fig. S8b). Furthermore, the time-resolved PL data also showed that more exposure leads to a decrease in both decay times and a smaller contribution of the fast-decaying component. Similar excitonic behaviors are induced by defects, thinner samples and higher temperatures.

Defects play a pivotal role in the nonradiative decay of Wannier excitons in semiconductors by serving as carrier-capturing centers for multiphonon emission and Auger recombination.[44] Because of the localized nature of the Cr-based excitons,[20] we propose that the defects in $CrPS_4$ provide a distinctive relaxation pathway by lowering the energy barrier for RISC, which is facilitated by the quartet-doublet coupling.[26] Structural defects, including vacancies and antisites, essentially lead to loose coordination of the central ions. Such defective LF found for $Cr^{3+}$ ions in glassy hosts is known to lower the energy of $^4T_{2g}$.[45] Because $^2E_g$ is hardly affected, the energy difference between the two excited states becomes smaller, facilitating RISC from $^2E_g$ to $^4T_{2g}$ and shortening the phosphorescence lifetime observed in Fig. 4. We note that the fluorescence is stronger for more defective bulk samples (CVT-grown) than flux-grown ones at 150 K, where both **F** and **P** bands are substantially populated (Fig. S4). Defect-induced **D** band is more prominent and strongly bound for 4L than bulk (Fig. S3). $QY_{eff}$ is smaller for samples with a larger contribution of the faster component (Fig. 4c). These facts support that RISC is more efficient for 4L because of increased defects than bulk samples. At high temperatures, defective LF also enhances the nonradiative decay of $^4T_{2g}$ to $^4A_{2g}$ and may quench some fraction of repopulated $^4T_{2g}$.[46] This explains the significantly reduced $QY_{eff}$ for < 7L (Fig. 4b).

The excitonic decay can be further analyzed based on an established kinetic model[24] for a three-level system consisting of $^4A_{2g}$, $^4T_{2g}$ and $^2E_g$ (see Supporting Note for a complete description). At the time (t) of zero, the two excited states are assumed to possess thermalized populations after an impulsive optical transition, $^4A_{2g} \rightarrow {}^4T_{2g}$. Whereas both excited states may relax to the ground state through radiative or nonradiative routes, they may also cross over to the other state via ISC or RISC, as depicted in Fig. S9. Then, the populations of the two states, [$^4T_{2g}$] and [$^2E_g$], can be described as biexponential functions of t. Because the latter is directly proportional to the intensity of **P** and approximately to the signal detected by the VIS CCD, the experimental $\tau_1$ and $\tau_2$ corresponds to the time constants for the population decay of [$^2E_g$]. In the limit of low RISC rate ($k_{RISC}$), $\tau_1$ decreases, and the contribution of the fast component increases with increasing $k_{RISC}$. This prediction corroborates our conclusion that the decrease in $\tau_1$ for thinner samples (Fig. 4a) and higher temperature (Fig. 3c) is due to increased $k_{RISC}$. It is also to be noted that $1/\tau_1$ ($1/\tau_2$) converges to $k_T$ ($k_E$), which is the sum of the rate constants for the transitions from $^4T_{2g}$ ($^2E_g$) (Supporting Note). On the other hand, the fact that $\tau_2$ also changed in the same direction as $\tau_1$ is not consistent with the model. This finding implies that the three-level model may be a overly simplified picture and a better description would require refinements including temperature-dependence of non-radiative decays. Alternatively, the observed biexponential decay may be attributed to the presence of two types of $Cr^{3+}$ ions with distinctive PL lifetimes. The contribution of this scenario would not be major because otherwise it would have to explain the temperature-dependent bifurcations in PL.

**Conclusion**

In summary, we unraveled the key relaxation steps of $Cr^{3+}$-localized excitons in 2D and bulk $CrPS_4$ using steady-state and time-resolved PL spectroscopy with NIR and VIS CCDs. With decreasing temperature from 298 to 4 K, a broad fluorescence band at 1.13 eV was replaced by a multi-peak phosphorescence band at 1.36 eV (< 100 K), some fraction of which was converted to a defect-mediated band at 1.08 eV (< 50 K). The temperature-dependent bifurcations in PL could be explained using a three-level kinetic model, where two excited states of different spin multiplicity convert to each other via the intersystem crossing. The competition between the two radiative transitions was also directly mapped from time-resolved PL measurements performed as a function of temperature, thickness and defect density. As branching into spin flip-derived electronic states is common in d-d transitions, similar PL bifurcations can be observed for other 2D crystals of transition metals under ligand-field coordination.

**Methods**

***Preparation and characterization of samples.*** Two types of bulk $CrPS_4$ crystals were used: the one synthesized by chemical vapor transport (CVT) method[29] and the other grown by self-flux method (2D Semiconductors). 2D $CrPS_4$ samples were prepared by mechanical exfoliation of the bulk crystals onto amorphous quartz (SPI, SuperSmooth) and $SiO_2$/Si substrates.[29, 34] Unless otherwise noted, the former was used to avoid optical artifacts induced in the latter by multiple reflections and subsequent interference. To minimize oxidation in the ambient conditions, prepared samples were stored in a dark vacuum desiccator sustained below 25 Torr. The sample thickness was determined by optical contrast and atomic force microscopy (AFM).[29, 34] Optical micrographs were recorded using an optical microscope (Nikon, LV100) connected to a CMOS camera. The red (blue)-channel optical contrast, defined as the fractional change in reflection with respect to that of bare $SiO_2$/Si (quartz) substrates, gave the best sensitivity and reliability in resolving thickness. Height and phase AFM images were obtained with a commercial unit (Park Systems, XE-70) in the non-contact mode. The nominal radius of silicon tips was 8 nm (MicroMasch, NSC-15).

***Protective encapsulation by ALD method.*** To prevent the ambient and photoinduced oxidation, 10-nm thick $Al_2O_3$ films were deposited on few-layer and bulk $CrPS_4$ samples using atomic layer deposition (ALD). Trimethylaluminum (TMA) and $H_2O$ vapor were used respectively as a precursor and an oxidant for deposition of $Al_2O_3$. Each cycle of ALD consisted of four steps: feeding TMA for 0.3 s, purging Ar gas for 15 s, feeding $H_2O$ for 0.2 s, and purging Ar gas for 15 s. When conducted at 100 °C, each cycle led to a growth of 0.084 nm. The base pressure of the reactor was ~1 Torr with the presence of $N_2$ carrier gas (99.9999%) at a flow rate of 50 mL/min. Before deposition, samples were rapidly annealed for 120 s at 100 °C under $N_2$ flow at 500 mL/min to minimize possible surface contaminants.

***Steady-state PL measurements in visible (VIS) range.*** Steady-state PL measurements were performed with two home-built micro-spectroscopy setups, one with a visible-range CCD and the other with an NIR-range CCD. The former setup, described elsewhere,[47] is run by 140-fs pulses of 80 MHz (1.77 eV unless otherwise noted) from a Ti:Sapphire laser (Coherent, Chameleon). The excitation beam was focused onto a 1.6-μm spot in FWHM with a 40X objective lens (numerical aperture = 0.60). Back-scattered PL signals collected by the same objective lens were guided into a Czerny-Turner spectrometer (Andor, Shamrock 303i) that was equipped with the VIS CCD camera (Andor, Newton). A 785 nm long-pass filter was used in front of the spectrometer to block the excitation laser beam. For temperature-dependent measurements, samples were placed inside a liquid-nitrogen cryostat (Linkam, HFS600E-PB4) and probed with another objective lens (50X, numerical aperture = 0.55). The polarization of the excitation beam was set parallel to the *b* axis of samples (Figure 1b) because of possible anisotropy in PL signals. To minimize unwanted photodamage, the average power was maintained below 0.2 mW.

The latter NIR setup[48] was used in obtaining undistorted PL spectra below 1.3 eV and reaching down to 4 K. The spot size of the excitation beam from a HeNe laser (1.96 eV) was ~3 μm in FWHM when focused with an objective lens (15X, NA = 0.28). Back-scattered PL signals collected by the same objective lens were guided into a spectrometer (Princeton Instruments, SP2300) and then the NIR CCD camera (Princeton Instruments, PyLoN-IR:1024-1.7) at Materials Imaging & Analysis Center of POSTECH. The temperature of samples was varied in the range of 4 ~ 298 K using a cryostat (Montana Instruments, Cryostation s50) operated with liquid helium. An 800 nm long-pass filter was placed in front of the spectrometer to block the excitation beam and stray light. We maintained the average power below 0.2 mW to avoid photodamage and the polarization of the excitation beam parallel to the *b* axis.

**Time-resolved PL measurements.** PL signals were processed with time-correlated single photon counting (TCSPC) method using a TCSPC device (PicoQuant, PicoHarp 300) connected to the VIS setup described above. The start and stop signals for the device were obtained with a photodiode (PicoQuant, TDA200) and a single-photon avalanche diode (SPAD, Micro Photon Devices, $PD-100-CTC), respectively. The temporal resolution defined as the FWHM of instrumental response function (IRF) was 50 ps at 1.77 eV. The PL signals in the wavelength range of 800 ~ 1000 nm were selected using a set of edge filters. The average power of the excitation beam was 9 μW.

ASSOCIATED CONTENT

**Supporting Information.**

Spectral quantum efficiency of the employed near-infrared (NIR) and visible (VIS) CCD cameras; Analysis of NIR PL spectra of bulk $CrPS_4$; Variable-temperature NIR PL spectra of 4L $CrPS_4$; Comparison between CVT and flux-grown bulk samples; Time-domain PL behavior of bulk $CrPS_4$; PL intensity transients of 2 ~ 250L $CrPS_4$; Effects of photon fluence on PL transients; Effects of UV-ozone (UVO) treatments; Kinetic model for a three-level system. This material is available free of charge via the Internet at http://pubs.acs.org.

AUTHOR INFORMATION

Corresponding Authors

*E-mail: jonghwankim@postech.ac.kr, sunryu@postech.ac.kr

#These authors contributed equally.

Notes

The authors declare no conflict of interest.

ACKNOWLEDGMENTS

The authors thank Myeongkee Park for assistance with the computer code to analyze PL transients. This work was supported by Samsung Science and Technology Foundation (Project Number SSTF-BA1702-08), National Research Foundation of Korea (NRF-2022R1A4A1033247),

POSCO Green Science Project, and Samsung Electronics Co., Ltd (IO201215-08191-01). J.K. and S.Y. acknowledge the support from the National Research Foundation of Korea grants (NRF-2020R1A4A1018935, and 2020R1A2C2103166). C.L. acknowledges the support from the National Research Foundation of Korea grants (NRF-2020R1A2C2014687). J. Kim, G. Jin and M.-H. Jo acknowledge the support from the Institute for Basic Science (IBS) under Project Code IBS-R034-D1.

**FIGURES & CAPTIONS**

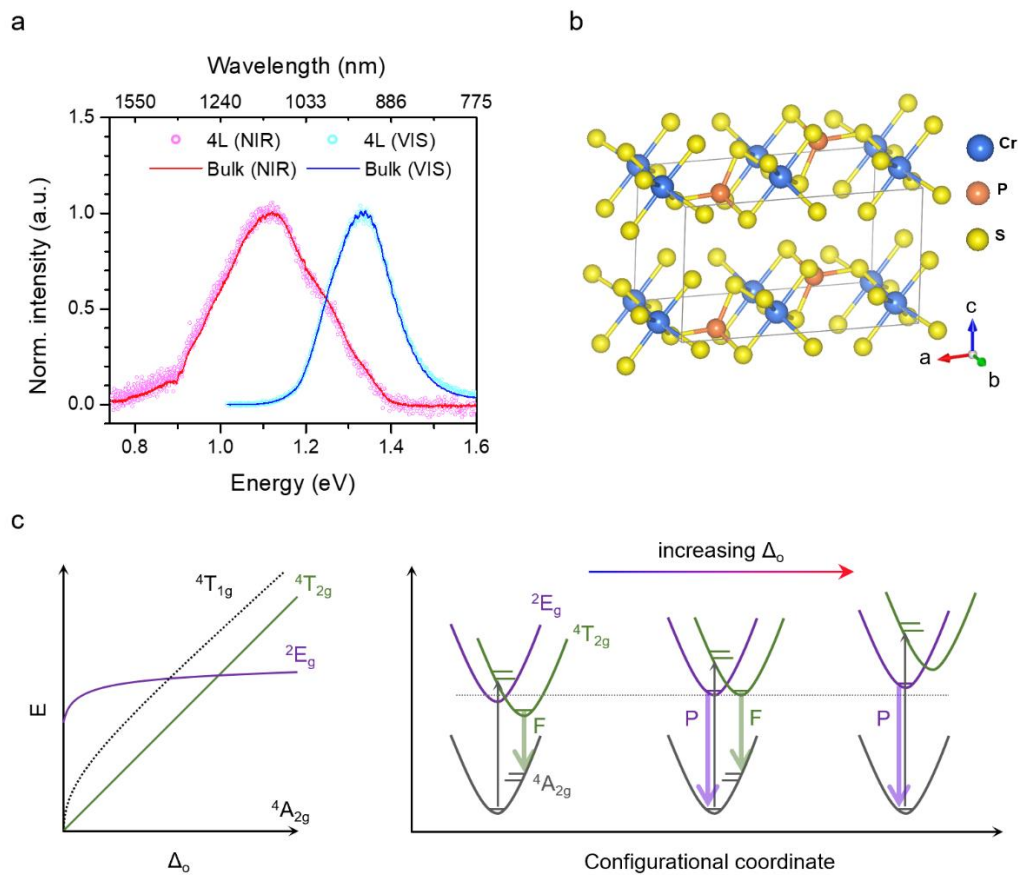

**Figure 1.** Dual-band NIR PL of CrPS$_4$ and its electronic origin. (a) PL spectra of 4L (circles) and bulk (lines) CrPS$_4$ obtained with VIS (blue) and NIR (red) CCDs at 298 K. (b) Crystal structure of 2L CrPS$_4$ with a unit cell (black line). (c) Tanabe-Sugano (left) and configuration coordinate (right) diagrams for d$^3$ Cr$^{3+}$ under octahedral ligand field (LF) with varying splitting energy $\Delta_o$. Olive and purple arrows represent fluorescence (**F**, $^4T_{2g} \rightarrow {}^4A_{2g}$) and phosphorescence (**P**, $^2E_g \rightarrow {}^4A_{2g}$), respectively, following optical absorption (black arrow, $^4A_{2g} \rightarrow {}^4T_{2g}$).

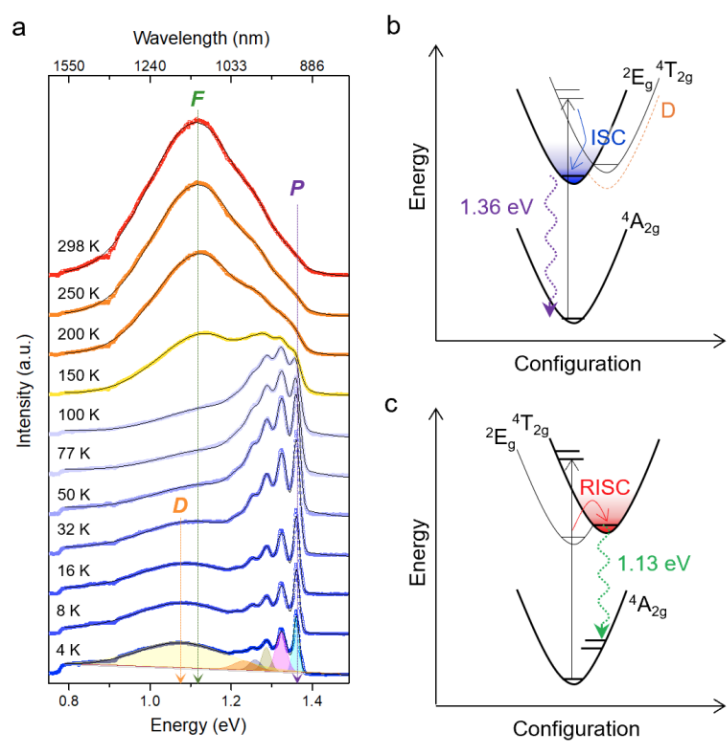

**Figure 2.** Temperature-dependent bifurcations of PL. (a) Variable-temperature PL spectra of bulk CrPS$_4$ obtained with an NIR CCD camera. Data (circles) fitted with Gaussian functions (solid lines) showed **F**, **P** and **D** bands, respectively, originating from fluorescence, phosphorescence and defect-mediated PL. Multiple color-shaded peaks of **P** correspond to a vibrational progression, 0-n (n = 0 ~ 4; from right to left). (b ~ c) Excitonic relaxation shown in the configuration coordinate diagrams for low (b) and high (c) T. Cr$^{3+}$ ions photoexcited to $^4T_{2g}$ or higher states undergo fast vibrational relaxation and intersystem crossing (ISC) to $^2E_g$. At low T, the doublet state then radiatively decays by generating **P** (purple arrow) at 1.36 eV. At high T, back-intersystem crossing (RISC) brings the ions back to $^4T_{2g}$ leading to delayed fluorescence (**F**; green arrow) at 1.13 eV. The dotted line below the $^2E_g$ minimum in (b) represents a shallow trap state responsible for **D**.
18

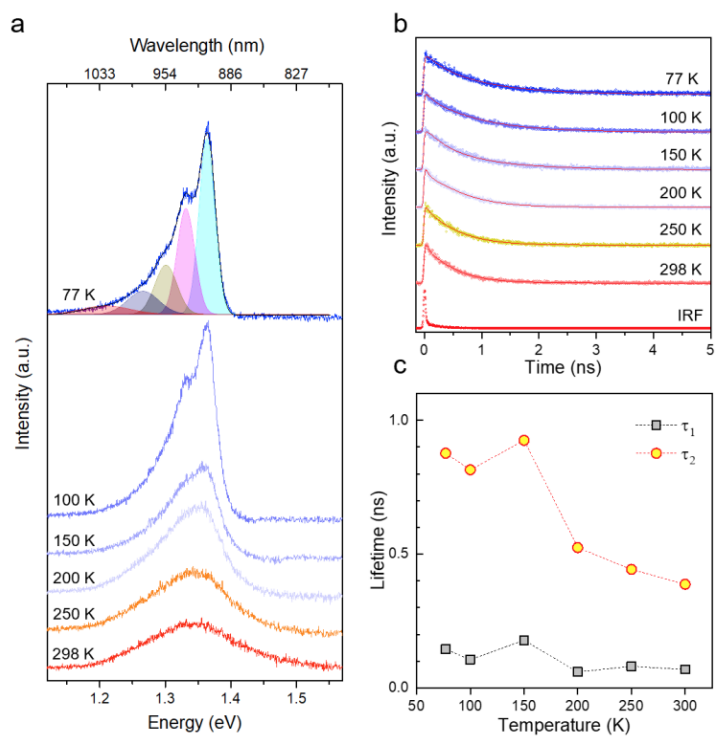

**Figure 3.** Time-domain verification of RISC. (a) Variable-temperature PL spectra of 6L $CrPS_4$ obtained with a VIS CCD. Despite the sensitivity attenuation of the detector below ~1.35 eV, the vibrational progression of **P** fitted with a multi-Gaussian function (solid line with color shades) is evident at 77 K. With increasing T, **P** is substantially reduced in intensity, losing the fine feature. (b) PL intensity transients of 6L obtained by time-correlated single-photon counting (TCSPC). Signals (circles) selected in 800 ~ 1000 nm were fitted with a biexponential function (solid lines) based on a three-level kinetic model (Supporting Note). The instrument response function (IRF) of the setup was numerically convoluted with the fit function. (c) Decay time constants ($\tau_1$ & $\tau_2$) extracted from (b).

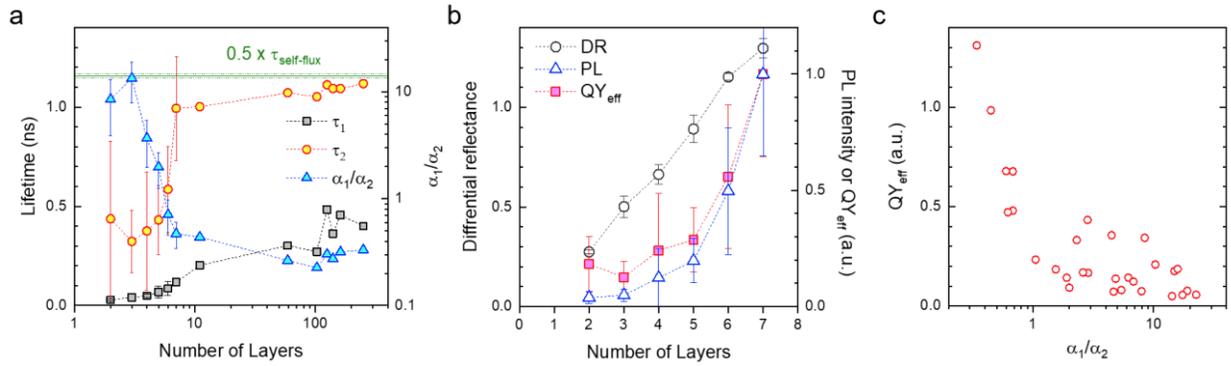

**Figure 4.** Enhanced PL decay in thinner $CrPS_4$. (a) Decay time constants ($\tau_1$ & $\tau_2$) and amplitude ratio ($\alpha_1/\alpha_2$) extracted from PL intensity transients of 2 ~ 250L obtained at 298 K (Fig. S6). The PL lifetime of flux-grown bulk samples is marked by the solid olive line with error bounds (olive dashed lines). (b) Differential reflectance (DR), PL intensity and effective PL quantum yield ($QY_{eff}$) of 2 ~ 11L. The latter two were normalized against 7L. (c) Relation between $QY_{eff}$ and $\alpha_1/\alpha_2$.

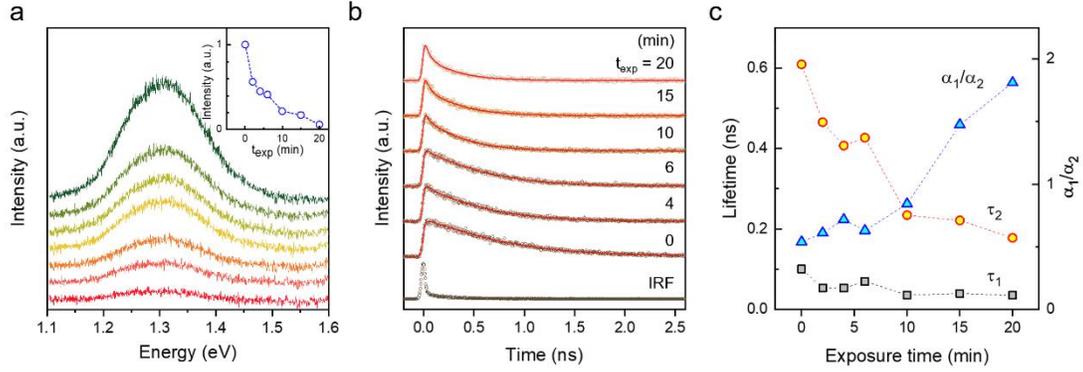

**Figure 5.** PL modulation by artificial defects. (a) PL spectra of 6L CrPS$_4$ obtained for various exposure time ($t_{exp}$) to UV-generated ozone: 0, 2, 4, 6, 10, 15 and 20 mins (top to bottom). The spectra were offset for clarity. The inset presents the PL intensity as a function of $t_{exp}$. (b) PL transients obtained as a function of $t_{exp}$. (c) Decay time constants ($\tau_1$ & $\tau_2$) and amplitude ratio ($\alpha_1/\alpha_2$) extracted from (b) as in Fig. 3.

TOC Figure

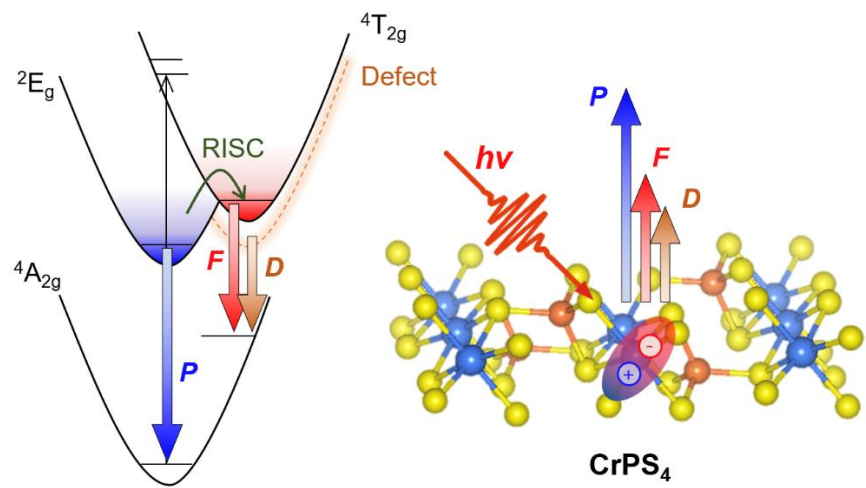